\documentclass[pra,twocolumn,preprintnumbers,amsmath,amssymb,showpacs,
nofootinbib,floatfix]{revtex4}

\usepackage{graphicx,bm,amsmath,latexsym}

\makeatletter
\def\graphicscale{\twocolumn@sw{0.3}{0.35}}
\def\graphicthreescale{\twocolumn@sw{0.3}{0.35}}

\usepackage{color}

\makeatother

\begin{document} \title{ Exact relations between particle fluctuations
and entanglement in Fermi gases }

\author{
Pasquale Calabrese, Mihail Mintchev and Ettore Vicari
}
\affiliation{Dipartimento di Fisica dell'Universit\`a di Pisa and INFN, 
Pisa, Italy}

\date{\today}

\begin{abstract}

We derive exact relations between the R\'enyi entanglement entropies
and the particle number fluctuations of (connected and disjoint)
spatial regions in systems of $N$ noninteracting fermions in arbitrary
dimension.  We prove that the asymptotic large-$N$ behavior of the
entanglement entropies is proportional to the variance of the particle
number.  We also consider 1D Fermi gases with a localized impurity,
where all particle cumulants contribute to the asymptotic large-$N$
behavior of the entanglement entropies. The particle cumulant
expansion turns out to be convergent for all integer-order R\'enyi
entropies (except for the von Neumann entropy) and the first few
cumulants provide already a good approximation.  Since the particle
cumulants are accessible to experiments, these relations may provide a
measure of entanglement in these systems.
  
\end{abstract}


\maketitle


The nature of the quantum correlations of many-body systems, and in
particular the entanglement phenomenon, are fundamental physical
issues. They have attracted much theoretical interest in the last few
decades, due to the impressive progress in the experimental activity
in atomic physics, quantum optics and nanoscience, which has provided
a great opportunity to investigate the interplay between quantum and
statistical behaviors in particle systems. The great ability in the
manipulation of cold atoms in optical lattice (see, e.g.,
Ref.~\cite{BDZ-08}) has allowed the realization of physical systems
which are accurately described by theoretical models such as Hubbard
and Bose-Hubbard models in different dimensions, achieving through
experimental checks of the fundamental theoretical paradigma of
condensed matter physics.

The quantum correlations arising in the ground state of quantum
many-body systems can be characterized by the expectation values of
the products of {\em local} operators, such as the particle density
and one-particle operators, or by their integral over a space region
$A$, such as the particle-number correlators within $A$,
\begin{equation}
\langle N_A^m \rangle_c = 
\int_A \prod_{i=1}^m d^dx_i \langle 
\prod_{i=1}^m n({\bf x}_i)\rangle_c,
\label{cumul}
\end{equation}
where $n({\bf x})$ is the particle-density operator and 
\begin{equation}
N_A = \int_A d^dx \,n({\bf x})
\label{nadef}
\end{equation}
counts the number of particles in $A$.  Quantum
correlations are also characterized by the fundamental phenomenon of
entanglement, which gives rise to nontrivial connections between
different parts of extended quantum systems~\cite{rev}.  A widely
accepted measure of entanglement is given by the R\'enyi entropies of
the reduced density matrix $\rho_A$ of a subsystem $A$:
\begin{equation}
S^{(\alpha)}_A = \frac{1}{1-\alpha} \ln {\rm Tr}\rho_A^\alpha,
\label{saldef}
\end{equation}
whose limit $\alpha\to 1$ provides the von Neumann (vN) entropy.
Local correlations and bipartite entanglement entropies provide
important and complementary information of the quantum features of
many-body systems, of their ground states and of their unitary
evolutions, because they probe different features of the quantum
dynamics.  However, the entanglement entropy is a highly {\it
nonlocal} quantity which is difficult to measure.  Designing an
experimental protocol for its measurement represents a major challenge.

A recent interesting proposal considers the particle fluctuations as
effective probes of many-body
entanglement~\cite{KRS-06,KL-09,SRL-10,SRFKL-11}.  This is based on
the result that, for non-interacting fermions, one can write down a
formal expansion of the entanglement entropies of bipartitions in
terms of the even cumulants $V^{(2k)}_A$ of the particle-number
distribution, which can be defined through a generator function
as
\begin{equation} 
V^{(m)}=(-i\partial_\lambda)^m 
\ln \langle e^{i\lambda N_A} \rangle |_{\lambda=0}.
\label{cumdef}
\end{equation}
Indeed, the R\'enyi entropies can be written as~\cite{KL-09,SRFKL-11}
\begin{eqnarray} 
S^{(\alpha)}_A&&= \sum_{k=1}^\infty s^{(\alpha)}_k V^{(2k)}_A,
\label{sv2ia} 
\\&&\!\!\!
s^{(\alpha)}_k= 
\frac{(-1)^k (2\pi)^{2k}2  \zeta[-2k,(1+\alpha)/2]}{(\alpha-1) \alpha^{2k} (2k)!}, 
\end{eqnarray}
where $\zeta$ is the generalized Riemann zeta function.
In particular, for the lowest integer $\alpha$ we have
\begin{eqnarray} 
&&S^{(1)}_A = {\pi^2\over 3} V^{(2)}_A 
+ {\pi^4\over 45} V^{(4)}_A 
+ {2\pi^6\over 945} V^{(6)}_A + ...
\label{s1vn} \\
&&S^{(2)}_A = {\pi^2\over 4}
V^{(2)}_A - {\pi^4\over 192} V^{(4)}_A + {\pi^6\over 23040} V^{(6)}_A
+ ...
\label{s2vn}
\end{eqnarray} 
The possibility of turning these expansions into an effective measure of
entanglement depends on its convergence properties, which appear
problematic due to the behavior of the coefficients $s^{(\alpha)}_k$
with increasing $k$. 

In this paper we investigate the relations between entanglement entropies
and particle fluctuations, and in particular the convergence properties
of the formal expansions (\ref{sv2ia}-\ref{s2vn}). We will show
that, in {\it noninteracting fermion
gases} with $N$ particles in a finite volume of any dimension $d$, the
expansion (\ref{sv2ia}) gets effectively truncated in the large-$N$
limit, because the high cumulants $V^{(n)}_A$ with $n>2$ are all
suppressed relatively to the particle variance 
\begin{equation}
V^{(2)}_A=\langle N_A^2 \rangle_c\equiv 
\langle N_A^2 \rangle -\langle N_A \rangle^2.
\label{v2na}
\end{equation}
The leading $N^{(d-1)/d} \ln
N$ asymptotic behavior of $S^{(\alpha)}_A$ in Eqs.~(\ref{sv2ia}),
(\ref{s1vn}) and (\ref{s2vn}) arises from $V^{(2)}_A$ only,
 because the leading order
of each cumulant $V^{(k)}$ with $k>2$ vanishes for any subsystem $A$
(including disjoint ones) in any dimension.  This implies the general
asymptotic relation
\begin{equation} 
\frac{S^{(\alpha)}_A}{V^{(2)}_A} =
\frac{(1+\alpha^{-1})\pi^2}{6} + o(1). \label{anyd} 
\end{equation}
We will also consider the effect of a localized impurity in 1D fermion
gases.  In this case all particle-number cumulants contribute to the
leading logarithmic behavior of the R\'enyi entropies. The expansion
(\ref{sv2ia})
turns out to be convergent for the integer R\'enyi entropies, so that
a few particle cumulants provide a good estimate, with the exception
of the vN entropy which appears problematic in this respect.
Relations like Eq.~(\ref{anyd}) may provide an experimental measure of
entanglement.

This letter is organized as follows. 
First we consider Fermi gases in any dimension, 
and prove  Eq.~(\ref{anyd}) for any subsystem $A$, by 
rigorously computing the asymptotic 
behaviors of  all cumulants of the
particle distribution.
Then we focus on one-dimensional Fermi gases,
for which more general results,
including subleading terms, can be obtained
for the asymptotic behaviors of 
the particle fluctuations and entanglement entropies
of connected and disjoint subsystems.
Finally, we extend our analysis to one-dimensional Fermi gases
in the presence of a defect.

{\bf The leading behavior in arbitrary dimension.}--
We consider a system of $N$ non-interacting spinless fermions with
discrete one-particle energy spectrum, which may arise from a finite
volume or an external potential.  The many-body ground state is
obtained by filling the lowest $N$ one-particle energy levels
$\phi_n(x)$. The cumulants $V^{(m)}_A$ of the particle-number
distribution and the entanglement entropies of a subsystem $A$ can be
written in terms of the overlap matrix~\cite{CMV-11,CMV-11-a}
\begin{equation}
{\mathbb A}_{nm} =  \int_A d^dz\, \phi_n^*({\bf z}) \phi_m({\bf z}),
\qquad n,m=1,...,N.
\label{aiodef}
\end{equation}
The eigenvalues $a_i$ of ${\mathbb A}$ are real and limited, $a_i \in (0,1)$. 
The matrix 
\begin{equation}
{\bar{\mathbb A}} = {\mathbb I} -{\mathbb A}
\label{comA}
\end{equation}
 is the overlap matrix of the complement of the region $A$.

The particle fluctuations within a region $A$ can be characterized 
by the cumulants $V^{(m)}$ of the particle distribution.
 The cumulant generator function for lattice free fermions has been derived 
in Ref. \cite{SRFKL-11} in terms of the two-point correlation function ${\mathbb C}_A$ 
restricted to the subsystem $A$. 
From this, taking the continuum limit and using the 
fundamental property
\begin{equation}
{\rm Tr}\, {\mathbb C}_A^n={\rm Tr}\, {\mathbb A}^n, \qquad {\rm for\, any}\; n\in{\mathbb N},
\end{equation} 
the cumulants generator function of the particle distribution can be expressed in terms
of the overlap matrix  ${\mathbb A}$ by replacing ${\mathbb C}_A$ with 
${\mathbb A}$, obtaining
\begin{eqnarray}
&&V^{(m)} = (-i\partial_\lambda)^m G(\lambda,{\mathbb A})|_{\lambda=0},\label{vny}\\
&&G(\lambda,{\mathbb A}) = {\rm Tr}\ln\left[1 + \left(e^{i\lambda} - 1\right) 
{\mathbb A}\right].
\label{ygen}
\end{eqnarray}
The even cumulants $V^{(2k)}$, which enter Eq.~(\ref{sv2ia}), 
can be cast in the form
\begin{eqnarray}
&&V^{(2k)}_A = 
\sum_{n=1}^k w_{k,n}  {\rm Tr}\,{\mathbb E}^n, \label{vegen}
\end{eqnarray}
where
\begin{eqnarray}
&&{\mathbb E} \equiv {\mathbb A} (1-{\mathbb A})= 
{\mathbb A} \bar{{\mathbb A}},\label{edef} \\ 
&&w_{k,n}=2\sum_{p=1}^n (-1)^{p+1} p^{2k} \frac{(2n-1)!}{(n-p)! (n+p)!}. 
\label{wkn}
\end{eqnarray}
In particular, 
\begin{equation}
V^{(2)}_A =  {\rm Tr}\,{\mathbb E},\quad
V^{(4)}_A = {\rm Tr}\,[{\mathbb E} - 6 {\mathbb E}^2].
\label{v24}
\end{equation}  
The entanglement entropies are obtained as~\cite{CMV-11,CMV-11-a}
\begin{equation}
S^{(\alpha)}_A =  
{1\over 1-\alpha} 
{\rm Tr} \ln[{\mathbb A}^\alpha + (1-{\mathbb A})^\alpha] .
\label{salai}
\end{equation}
In particular, the $\alpha=2$ entropy can be written as
\begin{eqnarray}
S^{(2)}_A = -{\rm Tr}\,\ln(1 - 2 {\mathbb E})=
\sum_{k=1}^\infty \frac{2^k}{k} {\rm Tr} \,{\mathbb E}^k.
\label{s2ek}
\end{eqnarray}
The eigenvalues of ${\mathbb E}$ satisfies $e_i\in (0,1/4)$, thus the
series (\ref{s2ek}) is convergent for any $N$, providing a systematic
approximation scheme in terms of $V^{(2k)}_A$ by inverting
Eq.~(\ref{vegen}).

In systems of noninteracting fermions and for arbitrary dimension $d$,
the entanglement entropy of connected bipartitions grows
asymptotically as $N^{(d-1)/d} \ln N$~\cite{CMV-11-c}. The logarithm
of the asymptotic behavior is related to the logarithmic area-law
violation in lattice free fermions~\cite{Wolf-06,GK-06,BCS-06,HLS-09,
DBYH-08,Sobolev-10,Swindle-10}.  In homogeneous systems with periodic
and open (hard-wall) boundary conditions (PBC and OBC respectively)
the prefactor can be analytically computed~\cite{CMV-11-c} using the
Widom conjecture~\cite{Widom-81}.  This method applies, and is even
more suited, to compute the large-$N$ behavior of particle
fluctuations with both PBC and OBC in any dimension.  In fact, unlike
the entanglement entropies, we deal with smooth functions, for which
the Widom conjecture has been proved~\cite{HLS-09,Sobolev-10}.  We
apply this theorem to the overlap matrix of a subsystem $A$ (with
smooth boundaries $\partial A$) of a finite system of size $L^d$ with
PBC, which is~\cite{CMV-11-c} 
\begin{equation}
{\mathbb A}_{nm} = L^{-d} \int_{A} d^d
x \; e^{i2\pi ({\bf k}_m-{\bf k}_n) \cdot {\bf x}/L }
\label{anmpa}
\end{equation}
with ${\bf k}\in {\mathbb Z}^d$ within the Fermi surface $\partial \Gamma$.  It
allows us to derive the large-$N$ behavior of ${\rm Tr}\,F({\mathbb
A})$, where $F(z)$ is any function analytic in $\{z\, :\, |z| <
1+\varepsilon \}$ with $F(0)=F(1)=0$, obtaining
\begin{eqnarray}
&&{\rm Tr}\,F({\mathbb A}) = C(F) N^{(d-1)/d} \ln N + o(N^{(d-1)/d}\ln N),
\nonumber \\
&&C(F) = {I(F)\over 4d\pi^{2}} 
\int_{\partial A} \int_{\partial f} dS_x 
dS_k |{\bf n}_{x} \cdot {\bf n}_{k}| ,
\quad\label{fff}
\end{eqnarray}
where we set $L=1$,
${\bf n}_x$ and ${\bf n}_k$ are the
normal vectors on $\partial A$ and on $\partial f$ which is
the Fermi surface $\partial\Gamma$ rescaled to enclose
a unit volume, and
\begin{equation}
I(F) = \int_0^1 d z \frac{F(z)}{z(z-1)}.
\label{IF}
\end{equation}
Note that the function $F$ enters only the integral $I(F)$.
The result (\ref{fff}) applies also to OBC.

The asymptotic large-$N$ behavior of ${\rm Tr}\,{\mathbb E}^{k}$ can
be computed using Eqs.~(\ref{fff}) and (\ref{IF}). 
Indeed, it corresponds to the function 
\begin{equation}
F_k(z) = z^k (1-z)^k,
\label{fkze}
\end{equation}
 thus
\begin{equation}
I(F_k) = \frac{[(k-1)!]^2}{(2k-1)!}. \label{fkz}
\end{equation}
 Plugging the last result into Eq. (\ref{fff}), we finally have
\begin{equation}
{\rm Tr}\,{\mathbb E}^k= N^{1-\frac{1}d} \ln N \frac{[(k-1)!]^2}{4d\pi^{2} (2k-1)! } \!
\int_{\partial A} \! \int_{\partial f} \! \! dS_x 
dS_k |{\bf n}_{x} \cdot {\bf n}_{k}| .
\label{trEk}
\end{equation}
The large-$N$ leading behaviors of $V^{(2k)}_A$  are obtained by
inserting these asymptotic results into their expressions in terms of
${\rm Tr}\,{\mathbb E}^k$, cf. Eq. (\ref{vegen}). 
  For any spatial region $A$ in any dimension $d$,  the variance
$V^{(2)}_A$ is
\begin{equation}
V^{(2)}_A= N^{1-{1}/d} \ln N \frac{1}{4d\pi^{2} } 
\int_{\partial A} \! \int_{\partial f}  dS_x 
dS_k |{\bf n}_{x} \cdot {\bf n}_{k}| .
\end{equation}
while, {\it very remarkably}, this leading term cancels for higher cumulants.  
 For odd cumulants $V^{(2k+1)}$ the leading term vanishes, because
they are odd under ${\mathbb A}\to {\mathbb I} - {\mathbb A}$.

 In the sum (\ref{sv2ia}), the leading behavior of the R\'enyi entropies 
gets a finite contribution only from the variance 
and the resulting entropies agree the direct computation  in Ref. \cite{CMV-11-c}.
Taking the ratio, the asymptotic large-$N$ relation (\ref{anyd}) follows. 
The above calculations do not allow us
to determine the behavior of the suppressed corrections in
Eq.~(\ref{anyd}).  Finite-$N$ calculations up to $N=O(10^3)$ 
indicate that they are $O(1/\ln N)$.  This is also supported by the analytic calculations in 1D
systems reported below.

Notice that Eq. (\ref{anyd}) relating particle fluctuations and
entanglement entropies can  also be obtained for lattice free fermions in
the thermodynamic limit, exploiting the
correspondence~\cite{CMV-11-a,CMV-11-c} between the overlap matrix
${\mathbb A}$ and the lattice two-point function ${\mathbb C}_{ij}$
where $i,j$ are the lattice sites within the region $A$.  
Indeed, analogous formulas for the particle-number cumulants hold by replacing
${\mathbb A}$ with ${\mathbb C}$ and the lattice version of some of the above 
results
has been already derived (as e.g. in Refs.~\cite{ELR-06,GK-06,SRFKL-11}).  
Moreover, developing the results of
Refs.~\cite{CV-10,CV-10-2}, analogous results can be also inferred for
free fermion gases in external potential, such as an harmonic one
which is usually present in experiments of cold atoms~\cite{BDZ-08}.

{\bf One-dimensional Fermi gas.}--
For 1D systems one may also consider an alternative computation based
on the Fisher-Hartwig conjecture~\cite{fh} and
generalizations~\cite{idk-09}, similarly to what has been done for the
entanglement entropies~\cite{jk-04,fc-11,CMV-11-a}.  This exact
approach allows us to calculate not only the leading term in
$V^{(m)}_A$ but also the leading and subleading corrections to the
scaling. 

Let us consider a 1D system of size $L$ with PBC or OBC, and
the interval $A=[0,x]$ as subsystem.  Setting $L=1$, the corresponding
overlap matrix is~\cite{CMV-11}
\begin{equation}
{\mathbb A_{nm}} =  {\mathbb P_{nm}} (x) \equiv   
{\sin[\pi(n-m)x] \over \pi(n-m)} 
\qquad ({\rm PBC})
\label{Snmpbc}
\end{equation}
and
\begin{eqnarray}
{\mathbb A_{nm}} &=& {\mathbb O_{nm}}(x)\equiv \label{Snmobc}\\ 
&&\equiv {\sin[\pi(n-m)x] \over \pi(n-m)} -  {\sin[\pi(n+m)x]
\over \pi(n+m)} \qquad \!\! ({\rm OBC}).
\nonumber
\end{eqnarray}
The asymptotic large-$N$
behavior of the entanglement entropies has been already reported in
Refs.~\cite{CMV-11,CMV-11-a}
\begin{equation}
S^{({\alpha})}(x)  ={1+\alpha^{-1}\over 6}
\ln(N\sin\pi x) + 
b_\alpha+ o(1)
\label{FHres2pbc}
\end{equation}
for PBC, and
\begin{equation}
S^{({\alpha})}(x)  ={1+\alpha^{-1}\over 12}
\ln(2N\sin\pi x) + 
\frac{b_\alpha}{2} + o(1)
\label{FHres2obc}
\end{equation}
for OBC, 
where also the constants $b_\alpha$ and the $o(1)$
corrections are known.   Fisher-Hartwig calculations can be  generalized to the
particle cumulants (details will be reported elsewhere), obtaining for
PBC
\begin{eqnarray}
{\rm Tr}\,{\mathbb E}^{k} & =&  
\frac1{\pi^2}\frac{[(k-1)!]^2}{(2k-1)!} \ln (2N\sin\pi x)+\label{v2k}\\
&+&i \int_{-\infty}^\infty dx 
\frac{g_k(\tanh\pi x)}{\cosh^2(\pi x)} 
\ln\frac{\Gamma(1/2 + i x)}{\Gamma(1/2 - i x)}+o(1)\,,
\nonumber
\end{eqnarray}
with 
\begin{equation}
g_k(z)=kz[(1-z^2)/4]^{k-1}/2.
\end{equation} 
The leading terms agree with Eq.~(\ref{trEk}), but
Eq.~(\ref{v2k}) also provides the $O(1)$ corrections.
Then, using Eq.~(\ref{vegen}), we obtain
\begin{eqnarray}
&&V^{(2)}(x) = {\rm Tr}\,{\mathbb E} =
{1\over \pi^2}\ln(N\sin\pi x) +{v_2} + o(1),\quad
\label{Vres2}\\
&&V^{(2k)}(x) = v_{2k} + o(1)\quad {\rm for}\;\;k>1,\label{Vres4}
\end{eqnarray}
where 
\begin{eqnarray}
&&v_2=(1+\gamma_E + \ln 2)/\pi^2,\label{v21d}\\
&&v_4=-0.0185104,\quad  v_6=0.00808937,
\label{v461d}
\end{eqnarray}
 etc.  The odd cumulants are suppressed,
\begin{equation}
V^{(2k+1)}(x) = o(1). 
\label{v2kp1}
\end{equation}
For OBC we find 
\begin{equation}
V^{(m)}_{\rm OBC}(N)=V^{(m)}_{\rm PBC}(2N)/2+o(1). 
\label{vobcpbc}
\end{equation}
 Following Ref.~\cite{fc-11},
the $o(1)$ subleading corrections can be systematically obtained.

{\bf Asymptotic behaviors for disjoint subsystems.}--
We now show that 
the asymptotic relation (\ref{anyd}) holds also for disjoint
subsystems, for which some exact results can be obtained.
Let us consider  a  Fermi gas of size $L=1$ with
 PBC.  The overlap matrix of a  generic disjoint subsystem
\begin{equation}
 [x_1,x_2] \cup [x_3,x_4],\quad 0<x_1<x_2<x_3<x_4<1,
\label{bregg}
\end{equation}
reads
\begin{eqnarray}
{\mathbb A}_{nm} &= &
e^{i\pi(n-m)(x_4+x_3)} {\mathbb P}_{nm}(x_4-x_3) \label{anm1234}\\
&+&e^{i\pi(n-m)(x_2+x_1)} {\mathbb P}_{nm}(x_2-x_1) ,
\nonumber
\end{eqnarray}
where the matrix
${\mathbb P}_{nm}$ is defined in Eq.~(\ref{Snmpbc}).
For  the particular subsystem
\begin{equation}
B = [0,x] \cup [1/2,1/2+x] , \quad 0<x<1/2,
\label{breg}
\end{equation}
the overlap matrix takes the simple form
\begin{equation}
{\mathbb A}_{nm} = 
e^{i\pi(n-m)\delta} [1 + (-1)^{n-m}]{\mathbb P}_{nm}(x) .
\label{anmb}
\end{equation}
Since we are interested in the traces of powers of ${\mathbb
A}_{nm}$, we may neglect the global phase. 
Then, exploiting the fact that all terms with different parity (odd $n-m$) vanish,
we have  for any integer $k$
\begin{equation}
{\rm Tr}\,{\mathbb A}^k_B(N) = 
2 {\rm Tr}\,{\mathbb A}^k_A(N/2),\quad A=[0,2x],
\label{exactreal}
\end{equation}
where ${\mathbb A}_A(N/2)$ is the overlap matrix of the
connected interval $A=[0,2x]$ with $N/2$ particles reported in Eq.~(\ref{Snmpbc}).
This exact relation implies
\begin{eqnarray}
&&S^{(\alpha)}_B(N)= 2 S^{(\alpha)}_A(N/2),\label{saldis}\\
 &&V^{(m)}_B(N) = 2 V^{(m)}_A(N/2). \label{v2dis}
\end{eqnarray} 
Using the asymptotic behaviors
(\ref{FHres2pbc}), (\ref{Vres2}) and (\ref{Vres4})
for the r.h.s.  of the above equations, 
we again recover  the asymptotic relation (\ref{anyd}).
Moreover,  we obtain the asymptotic behaviors of the
entanglement entropies
\begin{eqnarray}
S^{(\alpha)}_B  = 
{1+\alpha^{-1}\over 3}\left(
\ln N + \ln\sin 2\pi x\right) + b_\alpha + o(1),
\label{asybeh}
\end{eqnarray}
which agree with the free-fermion conformal field theory prediction \cite{cft-dis}
\begin{eqnarray}
&&S^{(\alpha)}_{[x_1,x_2] \cup [x_3,x_4]} = 
{1+\alpha^{-1}\over 6}
\Big[ \ln(4N^2) + \nonumber \\
&&
+ \ln{\sin(\pi x_{21})\sin(\pi x_{43})
\sin(\pi x_{41})\sin(\pi x_{32})\over \sin(\pi x_{31})\sin(\pi x_{42})}\Big]
\nonumber\\&&
+ b_\alpha + o(1), \label{cftpredis}
\end{eqnarray}
where $x_{ij}=x_i-x_j$ and
$0<x_1<x_2<x_3<x_4<1$.
While we explicitly reported the proof of  Eq. (\ref{anyd}) for the particular choice of the 
intervals in  Eq. (\ref{breg}), 
the proof can be extended to subsystems with an arbitrary union  of equal and equidistant
intervals. We expect that the validity of Eq. (\ref{anyd}) extends to the most general case of disjoint subsystems.

{\bf One-dimensional Fermi gas with an impurity.}--
We now investigate how the above relations between entanglement
entropy and particle fluctuations may change in the presence of
localized interactions, such as those arising from local impurities.
We consider 1D fermion gases with an impurity localized at the point
separating the system in two equal parts of size $L=1$, with hard-wall
boundary conditions at their ends.  
The allowed {\it scale invariant}
conditions at the vertex, describing the universal features arising
from the presence of a defect, are fully encoded in the scattering
matrix~\cite{w2}
\begin{equation} 
{\mathbb S}(T)= 
{1\over 1+\epsilon^2}\left(\begin{array}{cc}-1+\epsilon^2 & 2\epsilon\\ 
2\epsilon & 1-\epsilon^2  
\\ \end{array} \right), \quad T={2\epsilon\over 1+\epsilon^2},
\label{n21}
\end{equation} 
where $T$ is the transmission coefficient.  $T=1$ corresponds to full
transmission, i.e., no impurity, thus to the bipartition into two
equal parts of a 1D system with OBC.  The corresponding entropies and
cumulants can be obtained by setting $x=1/2$ in
Eqs.~(\ref{FHres2obc}-\ref{Vres4}). 

 For $|T|<1$ the ground-state
entanglement entropies and particle fluctuations of one of the two
edges can be again derived from its overlap matrix ${\mathbb A}$.  
For even $N$  we have~\cite{CMV-11,CMV-11-b}
\begin{eqnarray}
&&{\mathbb A}_{nm} = {2\epsilon\over 1+\epsilon^2}{\mathbb O}_{nm}(1/2) 
\quad {\rm for}\;n\ne m,
\label{bmn2}\\
&&{\mathbb A}_{nn} = {1\over 1+\epsilon^2} \;\; {\rm for \; odd\;}n,\quad 
{\mathbb A}_{nn} = {\epsilon^2 \over 1+\epsilon^2} \;\; 
{\rm for \; even\;}\,n,
\nonumber
\end{eqnarray}
where ${\mathbb O}_{nm}$ is defined in Eq.~(\ref{Snmobc}).
The symmetry $a_k\to 1-a_k$ of the spectrum of ${\mathbb A}$ implies that 
any odd observable with respect to ${\mathbb A} \to
1-{\mathbb A}$, such as the odd particle cumulants, vanishes.  

\begin{table}
\caption{
In order to check the
convergence of the expansions (\ref{sv2ia})  for the
coefficients $C_{S^{(\alpha)}}$ of the large-$N$ logarithmic behavior
$S^{(\alpha)}\approx C_{S^{(\alpha)}}\ln N$ in the presence of a defect
with $T=1/2$,
we report the sum $\sum_{k=1}^K s^{(\alpha)}_k C_{V^{(2k)}}$ for the
$\alpha=1,2,3$ Renyi entropies, whose leading-log coefficients are
$0.0570281,\; 0.0264623,\;0.0203204$ respectively.  An analogous
pattern of convergence is shown by the data at fixed $N$.}
\label{convcheck}
\begin{ruledtabular}
\begin{tabular}{clll}
\multicolumn{1}{c}{K}&
\multicolumn{1}{c}{vN ($\alpha=1$)}&
\multicolumn{1}{c}{$\alpha=2$}&
\multicolumn{1}{c}{$\alpha=3$}
\\
\colrule
1 & $\phantom{-}$0.0416667 & 0.0312500 & 0.0277778 \\
2 & $\phantom{-}$0.0622283 & 0.0264309 & 0.0201623 \\
3 & $\phantom{-}$0.0622283 & 0.0264309 & 0.0201623 \\
4 & $\phantom{-}$0.0288468 & 0.0264614 & 0.0203150 \\
5 & $\phantom{-}$0.0763883 & 0.0264626 & 0.0203256 \\
6 & $\phantom{-}$0.703467 & 0.0264623 & 0.0203209 \\
7 & $-$3.61878 & 0.0264623 & 0.0203202\\
8 & $-$47.3035 & 0.0264623 & 0.0203203 \\
9 & $\phantom{-}$949.44 & 0.0264623 & 0.0203204 \\
10 & $\phantom{-}$6860.34 & 0.0264623 & 0.0203204 \\
\end{tabular}
\end{ruledtabular}
\end{table}

Using the result of Ref. \cite{CMV-11-b},
\begin{eqnarray}
{\rm Tr}\,{\mathbb E}^k(T) = T^{2k}
{\rm Tr}\,{\mathbb E}^k(T=1),
\label{trent}
\end{eqnarray}
the particle fluctuations and entanglement entropies can be computed
using Eqs.~(\ref{vny}-\ref{s2ek}).  In particular,
for the particle variance we obtain
\begin{equation}
V^{(2)}(T) = {\rm Tr}\,{\mathbb E}(T)=T^2 V^{(2)} (T=1). 
\label{vtt1}
\end{equation}
This shows that, like the homogeneous case $T=1$, the particle
variance grows as $\ln N$, but with a smaller coefficient
$T^2/(2\pi^2)$.  Analogous results for the large-$N$ behavior of
higher cumulants follow and, unlike the homogeneous
case, also the higher-order even particle cumulants grow
logarithmically when $|T|<1$
\begin{equation}
V^{(2k)}\approx C_{V^{(2k)}}(T)\ln N.
\label{leadingv2k}
\end{equation}

Therefore, the relation (\ref{sv2ia})
between entanglement entropies and cumulants does not get truncated at
large $N$ and Eq. (\ref{anyd}) is not valid anymore.  
However, the large order behavior in Eqs.
(\ref{vegen}) and (\ref{sv2ia}) show that these infinite sums
are convergent for any {\it integer} $\alpha>1$, 
while for non-integer $\alpha$ they are {\it not} convergent.  
The truncated sum of
the leading large-$N$ logarithms of the particle cumulants
rapidly approaches the value of the
coefficient of the integer R\'enyi entropy, see Table~\ref{convcheck},
with the only exception of the vN entropy.  
In particular, using Eq.~(\ref{s2ek}), we obtain
\begin{eqnarray}
&&S^{(2)}  = C_{S^{(2)}} (T)\ln N+ O(1),\label{cs2t}\\
&& C_{S^{(2)}}  = \sum_{k=1}^\infty 
 \frac{2^{k-1}[(k-1)!]^2}{\pi^2 k (2k-1)!} T^{2k}=
{2\over \pi^2} [{\rm arcsin}(T/\sqrt{2})]^2.
\nonumber
\end{eqnarray}
The comparison with the entanglement entropy and particle variance of
two hard-core Bose-Hubbard (XX) chains of size $L$ separated by a bond
defect~\cite{EP-10,EG-10}, $S^{(2)}\approx C_{S^{(2)}}(T) \ln L$ and
$V^{(2)}\approx C_{V^{(2)}}(T) \ln L$ respectively, shows the
universality of the leading logarithmic behaviors.

{\bf Conclusions.}--

We showed that,
in a noninteracting Fermi gas with a large number $N$
of particles and in any dimension,  the R\'enyi
entanglement entropies of a spatial (connected or disjoint)
region $A$ are proportional to the
variance of the particle number in the same region, cf. Eq.~(\ref{anyd}),
with a coefficient that
does not depend  either on $A$ or on the space dimension of the system.
This remarkable result  is proved by first computing 
the asymptotic behaviors of all particle cumulants $V^{(m)}$,
which are such that the leading order
of each cumulant $V^{(k)}$ with $k>2$ vanishes for any subsystem $A$
(including disjoint ones) in any dimension.  
Then,  the leading $N^{(d-1)/d} \ln N$ asymptotic behavior of 
$S^{(\alpha)}_A$  in the formal expansion 
$S^{(\alpha)}_A= \sum_{k=1}^\infty s^{(\alpha)}_k V^{(2k)}_A$
 arises from $V^{(2)}_A$ only.
Therefore, entanglement entropies 
are directly related to the integral of the density-density
correlation function which is accessible in cold atoms experiments
through Bragg spectroscopy~\cite{bragg}, thus providing an
experimental measurement of the zero-temperature entanglement.
Furthermore the recent developed technique of single atom-site 
imaging technique \cite{bakr-10} should allow to measure all the lowest 
particle-number cumulants.

The situation is more involved for interacting systems.  Already, in systems
with localized interactions arising from impurities, all the cumulants
$V^{(2k)}$
contribute to the asymptotic large-$N$ behavior of the entanglement
entropies in the expansion 
$S^{(\alpha)}_A= \sum_{k=1}^\infty s^{(\alpha)}_k V^{(2k)}_A$.
 However, the resulting sum turns out to be rapidly
convergent for all integer R\'enyi entropies (except for the vN one),
and a few cumulants are enough to get a precise estimate of the
entanglement.  

The conservation of a global charge, and in particular
the particle number, is crucial for the connections between bipartite
entanglement and particle fluctuations. For interacting systems not
conserving the particle number, the entanglement should be related to
the more fundamental energy transport. Recent proposals
\cite{hgf-09,c-11,KL-09} to measure the entanglement entropy in
general 1D systems have considered protocols based on appropriate
quantum quenches.

Finally, let us mention that the relations between particle
fluctuations and entanglement entropies apply also to 1D Bose gases
with short-ranged repulsive interactions.  Indeed, in the limit of
strong interaction (i.e. a gas of hard-core bosons), this can be
mapped to free fermions, so that their particle-number cumulants and
entanglement entropies coincide for connected regions.  The hard-core
Bose gas also describes the dilute limit of the finite-strength
models~\cite{LL-63}, which implies that it also provides their
infinite size limit at fixed $N$~\cite{CV-10-2} (with $O(L^{-1})$
corrections).  Therefore, in this regime, the same asymptotic
large-$N$ behavior is expected, and in particular Eq.~(\ref{anyd})
remains valid.

{\it Acknowledgements}. 
We thank C.~Flindt and I.~Peschel for correspondence.
PC research was supported by ERC under the
Starting Grant n. 279391 EDEQS.

\end{document}